\documentclass[lettersize,journal]{IEEEtran}

\usepackage{graphicx,multicol}
\graphicspath{{Figures/}}
\usepackage{psfrag,amsbsy,pstricks,amsmath,amsfonts,amssymb,cite}
\usepackage{algpseudocode}
\usepackage{epstopdf}
\usepackage{bm}
\usepackage{booktabs}
\usepackage[font=footnotesize]{subcaption}
\usepackage[font=footnotesize]{caption}
\usepackage{glossaries}
\usepackage{float}
\usepackage{hyperref}
\newacronym{AD}{AD}{angle-delay}
\newacronym{ADD}{ADD}{angle-delay domain}
\newacronym{AFD}{AFD}{angle-frequency domain}
\newacronym{AS}{AS}{activity state}
\newacronym{AWGN}{AWGN}{additive white Gaussian noise}

\newacronym{BS}{BS}{base station}
\newacronym{BICM-OFDM}{BICM-OFDM}{bit-interleaved coded \gls{OFDM}}
\newacronym{BICM}{BICM}{bit-interleaved coded}
\newacronym{BG}{BG}{Bernoulli Gaussian}
\newacronym{BG-MC}{BG-MC}{Bernoulli Gaussian-Markov chain}
\newacronym{B-TSGM}{B-TSGM}{Bernoulli two-state Gaussian mixture}

\newacronym{CDMA}{CDMA}{code division multiple access}
\newacronym{CE}{CE}{channel estimation}
\newacronym{CP}{CP}{cyclic prefix}
\newacronym{CS}{CS}{compressed sensing}
\newacronym{CSI}{CSI}{channel state information}

\newacronym{DD}{DD}{data detection}
\newacronym{DFT}{DFT}{discrete Fourier transform}

\newacronym{EM}{EM}{expectation maximization}

\newacronym{FDD}{FDD}{frequency-division duplexing}
\newacronym{FG}{FG}{factor graph}

\newacronym{GAMP}{GAMP}{generalized approximate message passing}
\newacronym{GAMP-BG-MC}{GAMP-BG-MC}{GAMP Bernoulli Gaussian-Markov chain}
\newacronym{GAMP-SBL}{GAMP-SBL}{GAMP sparse Basianyes learing}
\newacronym{GAMP-PCSBL}{GAMP-PCSBL}{GAMP pattern coupled sparse Basianyes learing}
\newacronym{GB}{GB}{grant-based}
\newacronym{GF}{GF}{grant-free}
\newacronym{GF-NOMA}{GF-NOMA}{grant-free non-orthogonal multiple access}

\newacronym{HMP}{HMP}{hybrid message passing}

\newacronym{ICI}{ICI}{inter-carrier interference}
\newacronym{IoT}{IoT}{Internet of Things}
\newacronym{ISI}{ISI}{inter-symbol interference}

\newacronym{LMMSE}{LMMSE}{linear minimum mean square error}

\newacronym{MC}{MC}{Markov chain}
\newacronym{MC-CDMA}{MC-CDMA}{multi-carrie code division multiple access}
\newacronym{M-GAMP}{M-GAMP}{multi-GAMP}
\newacronym{MIMO}{MIMO}{multiple-input multiple-output}
\newacronym{MIMO-OFDM}{MIMO-OFDM}{multiple-input multiple-output orthogonal frequency-division multiplexing}
\newacronym{mMTC}{mMTC}{massive machine-type communications}
\newacronym{MMV}{MMV}{multiple measurement vector}
\newacronym{MP}{MP}{message passing}
\newacronym{MSE}{MSE}{mean square error}

\newacronym{NMSE}{NMSE}{normalized mean square error}
\newacronym{NNSPL}{NNSPL}{nearest neighbor sparsity pattern learning}
\newacronym{NOMA}{NOMA}{non-orthogonal multiple access}

\newacronym{OFDM}{OFDM}{orthogonal frequency-division multiplexing}

\newacronym{PCSBL}{PCSBL}{pattern coupled sparse Basianyes learing}
\newacronym{PDF}{PDF}{probability density function}
\newacronym{PDFT-RP}{PDFT-RP}{partial DFT random permutation}

\newacronym{RA}{RA}{random access}

\newacronym{SBL}{SBL}{sparse Basianyes learing}
\newacronym{SCM}{SCM}{spatial channel model}
\newacronym{SER}{SER}{symbol error rate}
\newacronym{SFD}{SFD}{space-frequency domain}
\newacronym{STF}{STF}{structured Turbo framework}
\newacronym{SMV}{SMV}{single measurement vector}
\newacronym{SNR}{SNR}{signal-to-noise ratio}
\newacronym{STCS}{STCS}{structured turbo-compressed sensing}
\newacronym{STCS-DS}{STCS-DS}{STCS-delay support}
\newacronym{STCS-FS}{STCS-FS}{structured turbo-compressed sensing frequency support}

\newacronym{TSGM}{TSGM}{two-state
Gaussian mixture}

\newacronym{UE}{UE}{user equipment}
\newacronym{ULA}{ULA}{uniform linear array}

\newacronym{5G}{5G}{5th-generation}
\newacronym{6G}{6G}{6th-generation}

\hypersetup{pdfborder = {0 0 0}}
\usepackage{array}

\usepackage{textcomp}
\usepackage{stfloats}
\usepackage{url}
\usepackage{verbatim}
\usepackage{cite}
\usepackage[linesnumbered,ruled,vlined]{algorithm2e}
\SetKwInput{KwInput}{Input}                
\SetKwInput{KwOutput}{Output}              
\SetCommentSty{mycommfont}
\newcommand{\mr}[1]{\mathrm{#1}}
\newcommand{\bs}{\boldsymbol}

\begin{document}

\title{Hybrid Message Passing-Based Detectors for Uplink Grant-Free NOMA Systems}
\author{Yi Song, Yiwen Zhu, Kun Chen-Hu, \textit{Member, IEEE}, Xinhua Lu, Peng Sun*, Zhongyong Wang
\thanks{
Y. Song, Y. Zhu, P. Sun and Z. Wang are with the School of Electrical and Information Engineering, Zhengzhou University, 450001, Zhengzhou, China (songyizzu@gs.zzu.edu.cn; zhuyw@gs.zzu.edu.cn; iepengsun@zzu.edu.cn;iezywang@zzu.edu.cn). K. Chen-Hu is with the Department of Electronic Systems, Aalborg University, 9220, Aalborg, Denmark (kchenhu@es.aau.dk). X. Lu is with the College of Information Engineering, Nanyang Institute of Technology, 473000, Nanyang, China (ieluxinhua@sina.com).}
}
\maketitle

\begin{abstract}
This paper studies improving the detector performance which considers the \gls{AS} temporal correlation of the \glspl{UE} in the time domain under the uplink \gls{GF-NOMA} system. The \gls{BG-MC} probability model is used for exploiting both the sparsity and slow change characteristic of the \gls{AS} of the \gls{UE}. The \gls{GAMP-BG-MC} algorithm is proposed to improve the detector performance, which can utilize the bidirectional message passing between the neighboring time slots to fully exploit the temporally-correlated \gls{AS} of the \gls{UE}. Furthermore, the parameters of the \gls{BG-MC} model can be updated adaptively during the estimation procedure with unknown system statistics. Simulation results show that the proposed algorithm can improve the detection accuracy compared with the existing methods while keeping the same order complexity.
\end{abstract}


\section{Introduction}
\label{Sec:intro}
\IEEEPARstart{T}{he} \Gls{mMTC} is one of the key pillars of the next \gls{6G} cellular networks for \gls{IoT} applications \cite{Derrick5G}. In \gls{mMTC} scenario, the central \gls{BS} is serving billions of \glspl{UE}, and both spectral efficiency and low-latency must be satisfied by a multiple-access scheme. The \gls{GF-NOMA} is a promising candidate to support \gls{mMTC}. The NOMA technique enables the system resource block to simultaneously serve multiple \glspl{UE} \cite{DaiNOMA}, while the \gls{GF} scheme allows \glspl{UE} to transmit data to the \gls{BS} directly without waiting for its permission, which can decrease the access delay and reduce the control overhead for coordination \cite{LiuLiang}.

In the case of applying \gls{GF-NOMA} to uplink, recovering the transmitted signal through designing efficient multi-\glspl{UE} detection methods is challenging, not only due to the large number of the \glspl{UE} present in the cell, but they are also randomly accessing to the network in each access slot. 
However, \glspl{UE} are only sporadically and randomly activated in the practical \gls{mMTC} systems, which brings sparsity of the transmitted slots in the time-frequency resource grid \cite{LiBoslotCS}, and this sparsity can be changed across different slots within a single frame \cite{LiBoslotNOMA,GuoFTN-NOMA}.
Furthermore, once a \gls{UE} is activated for performing the data transmission, it typically requires several consecutive time slots. Then, this \gls{UE} will be deactivated for a certain period till the next transmission. This activation and deactivation is denoted as \gls{AS}, and it exhibits a temporal correlation~\cite{ZhangPCSBLTWC}.
Consequently, to enhance the performance of the detectors based on a \gls{GF-NOMA} system, not only the joint exploitation of sparsity but also the temporal correlation information provided by \gls{AS} in the system.

According to the literature of \gls{GF-NOMA}, detectors based on \gls{MP} algorithms \cite{GuoFTN-NOMA, ZhangPCSBLTWC,ZhangPCSBLTVT} have a better performance as compared to the conventional \gls{CS} methods. In \cite{GuoFTN-NOMA}, the authors exploited the temporal correlation of the active \gls{UE} and modeled by a Markov process. 
The authors in \cite{ZhangPCSBLTWC} utilized the \gls{SBL} and \gls{PCSBL} model to formulate the slow change of the \glspl{UE} \gls{AS}, then proposed the SBL-based multi-\glspl{UE} detection algorithms outperforming \gls{CS} methods.
Then \cite{ZhangPCSBLTVT} further embedded the \gls{GAMP} algorithm with the \gls{SBL}, \gls{PCSBL} models, and proposed the \gls{GAMP-SBL} and the \gls{GAMP-PCSBL} detector algorithms, which can reduce the complexity. However, the \gls{SBL} and \gls{PCSBL} models mainly consider the sparse structure of the \gls{UE} activity. Motivated by \cite{GuoFTN-NOMA}, we find that the \gls{MC} better captures the slow change of the \gls{AS} features with successive time slots and the active \gls{UE} sparsity simultaneously. Further different with \cite{GuoFTN-NOMA}, we use the \gls{HMP} rule~\cite{HMP-TSGM} to update the parameters of the \gls{MC} on the \gls{FG}, to improve the convergence speed.

In this paper, we focus on designing a high-performance \gls{MP} algorithm and resulting in the \gls{GAMP-BG-MC} detector algorithm for the uplink \gls{GF-NOMA} system. The main contributions of this paper are summarized as follows. 
\begin{enumerate}
    \item  We employ the \gls{BG-MC} model \cite{STCSFS} to fully exploit both the sparse structure of the active \gls{UE} and the slow change feature in consecutive time slots of the \gls{AS}. Indeed, we can improve the detector performance by updating the \gls{AS} by considering the temporal correlation. Hence, both the forward messages from the previous slot and the backward messages from the next slot are considered in the \gls{MP} algorithm at each iteration.
    \item Making use of the \gls{HMP} rule, we proposed the \gls{GAMP-BG-MC} algorithm to detect the received signal of multiple superimposed \glspl{UE}. The proposed \gls{GAMP-BG-MC} algorithm not only calculates the more accurate posterior \gls{PDF} of the transmitted signal but also adaptively learns the parameters of the prior model during the estimation procedure. Simulation results show that the \gls{GAMP-BG-MC} algorithm has a better \gls{SNR} performance as compared to the \gls{GAMP-SBL}~\cite{ZhangPCSBLTWC,ZhangPCSBLTVT} and the \gls{GAMP-PCSBL}~\cite{ZhangPCSBLTVT} algorithms respectively while keeping the complexity. 
\end{enumerate}

\section{System Model and Factor Graph}
\label{sec:system_model}
\subsection{System Model}
\begin{figure}[tbh]
    \centering
    \includegraphics[scale=0.6]{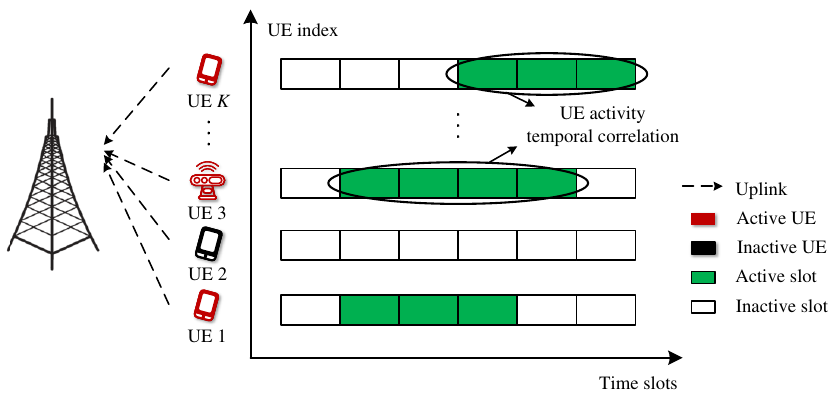}
    \caption{The left part of the figure denotes the uplink transmission from \glspl{UE} to the \gls{BS}; The right part of the figure indicates the sparse and temporally correlated structures of the transmit signal $\bs{B}$.}
    \label{fig:system_model}
\end{figure}

We consider a typical \gls{GF-NOMA} system \cite{ZhangPCSBLTVT}, where $K$ potential single-antenna \glspl{UE} share the same bandwidth resource and transmit the signal to the central \gls{BS} is equipped with a single antenna, as illustrated in Fig.~\ref{fig:system_model}. We assume that the transmission utilizes the \gls{MC-CDMA} scheme detailed in \cite{Fri2019MCCDMA}, which belongs to the \gls{NOMA} techniques. The \gls{MC-CDMA} is a hybrid scheme combing \gls{CDMA} and \gls{OFDM}, makes the \gls{MC-CDMA} signal tolerant against frequency selective channels, \gls{ISI} and \gls{ICI} through the use of cyclic prefix and equalization \cite{MC-CDMAagainstISI}. The uplink transmission of the \gls{UE}s is multiplexed over $N$ frequency-divided subcarriers. In each time slot, we assume the \gls{AS} of different \gls{UE}s are independent of each other, $K_a \le K$ of which are active and transmit packets to the \gls{BS}. We use the active rate $p_a$ to indicate the percentage of active \gls{UE}s. The active \gls{UE} $k$ sends the complex symbol $b_k$ from a predefined complex signal constellation, and spreads it over the $N$ sub-carriers. While the inactive \gls{UE} is considered transmitting a zero symbol $b_k = 0$. The received signal $\bs{y}=[y_1,\cdots,y_N]^{\mr{T}}$ at the \gls{BS} for the $j$-th time slot can be expressed as 
\begin{equation}
    \label{eq:SMV}
    \bs{y} = \bs{H}\bs{b}+\bs{w},
\end{equation}
where $\bs{H} = [\bs{h_1},\cdots,\bs{h_K}]\in \mathbb{C}^{N\times K}$ denotes the complex channel response matrix, representing the spread codes of $K$ \glspl{UE} over $N$ sub-carriers. $\bs{b} =[b_1,\cdots,b_K]^{\mr{T}}$ is the transmitted signal from $K$ \glspl{UE}. $\bs{w} =[w_1,\cdots,w_N]^{\mr{T}}\sim\mathcal{CN}(\bs{0},\lambda^{-1} \bs{I})$ denotes the \gls{AWGN} with the variance $\lambda^{-1}$. The received signals $\bs{Y}=[\bs{y}_1,\cdots,\bs{y}_J]\in \mathbb{C}^{N\times J}$ collect $J$ continuous time slots and represent as a \gls{MMV} model \cite{ZhangPCSBLTVT}
\begin{equation}
    \label{eq:MMV}
    \bs{Y} = \bs{H}\bs{B}+\bs{W},
\end{equation}
where $\bs{B} = [\bs{b}_1,\cdots,\bs{b}_J]\in \mathbb{C}^{K\times J}$ denotes the transmitted signal matrix, and the noise is $\bs{W} = [\bs{w}_1,\cdots,\bs{w}_J]\in \mathbb{C}^{N\times J}$. In a practical \gls{mMTC} scenario, the potential \glspl{UE} keep their \gls{AS} in some random consecutive time slots within the entire frame \cite{LiBoslotCS}. As shown in Fig.~\ref{fig:system_model}, the \gls{AS} of the \gls{UE} exhibits the temporal correlation in the time domain, which promotes the sparse structure of signal $\bs{B}$ in \eqref{eq:MMV}. The task of multi-\glspl{UE} detection is to estimate the sparse transmitted signal $\bs{B}$ from the received signal $\bs{Y}$ with the unknown \gls{UE} activity value.
\vspace{-0.1cm}

\subsection{Probabilistic Formulation}
From \eqref{eq:MMV}, we define the hidden variable $\bs{Z} \triangleq \bs{H}\bs{B} \in \mathbb{C}^{N\times J}$, in which $\bs{z}_j = \bs{H}\bs{b}_j,\forall j\in J$. Then the detector transfer function can be written as
\begin{equation}
    \label{eq:ChanTrans}
    p(\bs{Y}|\bs{B}) =  p(\bs{Y}|\bs{Z})p(\bs{Z}|\bs{B}) = {\textstyle \prod_{j=1}^{J}}p(\bs{y}_j|\bs{z}_j)p(\bs{z}_j|\bs{b}_{j}).
\end{equation}
The conditional \gls{PDF} $p(\bs{y}_j|\bs{z}_j)$ can be further factorized as
\begin{equation}
    \label{eq:Likelihood}
    p(\bs{y}_j|\bs{z}_j) = \prod_{n=1}^{N}\underset{f_{Y,n,j}}{\underbrace{p(y_{n,j}|z_{n,j})}} = \prod_{n=1}^{N}\mathcal{CN}(y_{n,j};z_{n,j},\lambda^{-1}).
\end{equation}
The hard constraint factors provided by $p(\bs{z}_j|\bs{b}_{j})$ can be concretely represented as 
\begin{equation}
    \label{eq:HardCons}
    p(\bs{z}_j|\bs{b}_{j}) = \prod_{n=1}^{N}\underset{f_{Z,n,j}}{\underbrace{p(z_{n,j}|\bs{b}_j)}} = \prod_{n=1}^{N} \delta\Big(z_{n,j}-\sum_{k=1}^{K}h_{n,k}b_{k,j}\Big),
\end{equation}
where the $h_{n,k}$ denotes the element in the $n$-th row and $k$-th column of $\bs{H}$.

For the channel prior model, we utilize the \gls{BG-MC} probability model \cite{STCSFS}, which can exploit both the block sparse structure and slow change of the \gls{UE} \gls{AS} feature simultaneously. Each element $b_{k,j}$ has a conditionally independent distribution given by
\begin{equation}
    \label{eq:ChaPrior}
    \begin{split}
         &p(\bs{B}|\bs{S},\bs{V}) = {\textstyle \prod_{k=1}^{K}}{\textstyle \prod_{j=1}^{J}}\underset{f_{B,k,j}}{\underbrace{p(b_{j,k}\,|\,s_{k,j},v_{k,j})}} 
         \\
         &= \delta (s_{k,j})\delta (b_{k,j}) +\delta (1-s_{k,j})\mathcal{CN}(b_{k,j};0,v_{k,j}^{-1}) ,
    \end{split}
\end{equation}
where $s_{k,j}\in \{0,1\}$ denotes the \gls{MC} state implicating if the \gls{UE} is active ($s_{k,j} = 1$) or inactive ($s_{k,j} = 0$) at the $j$ time slot; $v_{k,j}^{-1}$ is non-negative hyperparameter controlling the sparsity of the transmitted signal $b_{k,j}$. 

Similar to \cite{GuoFTN-NOMA,ZhuRAflame,STCSFS}, we employ the stationary first-order \gls{MC} to effectively capture the slow evolution of \gls{UE} \gls{AS} across time slots. The \gls{MC} here given by 
\begin{equation}
    \label{eq:Markovchain}
    p(\bs{S})=\prod_{k=1}^{K}p(\bs{s}_k)= \prod_{k=1}^{K}\big [ \underset{f_{S,k,1}}{\underbrace{p(s_{k,1})}} \prod_{j=2}^{J}\underset{f_{S,k,j}}{\underbrace{p(s_{k,j}|s_{k,j-1})}} \big ],
\end{equation}
with the initial probability given by $p(s_{k,1}) =(p^{10}_{k})^{s_{k,1}}(1-p^{10}_{k})^{1-s_{k,1}}$ and the transition probability is expressed as
\begin{equation}
\label{eq:Transition}
    p(s_{k,j}|s_{k,j-1}) =\left\{\begin{matrix}(1-p^{10}_{k})^{1-s_{k,j}}(p^{10}_{k})^{s_{k,j}},s_{k,j-1}=0, 
    \\
    (p^{01}_{k})^{1-s_{k,j}}(1-p^{01}_{k})^{s_{k,j}},s_{k,j-1}=1,\end{matrix}\right.
\end{equation}

The \gls{MC} is characterized by parameters $p^{01}_{k}\triangleq \mr{Pr}(s_{k,j}=0|s_{k,j-1}=1)$ and $p^{10}_{k}\triangleq \mr{Pr}(s_{k,j}=1|s_{k,j-1}=0)$ for different \glspl{UE}. Different from \cite{GuoFTN-NOMA,STCSFS}, we consider the parameters $p^{01}_{k},p^{10}_{k}$ as variables, and the prior given by the Beta distributions as 
\begin{equation}
    \label{eq:ProbabilityBeta}
    \underset{f_{p^{01}_k}}{\underbrace{p(p^{01}_{k})}} =  \mr{Be}(p^{01}_{k};c_{k},d_{k}),
    \underset{f_{p^{10}_k}}{\underbrace{p(p^{10}_{k})}} =  \mr{Be}(p^{10}_{k};e_{k},f_{k}),
\end{equation}
where $c_{k}, d_{k}$ and $e_{k}, f_{k}$ can control the initial values of $p^{01}_{k}$ and $p^{10}_{k}$, respectively. Meanwhile, the precision terms $v_{k,j}$ can be modeled as the Gamma distribution~\cite{Carles}, i.e.,
\begin{equation}
    \label{eq:PrecisionGamma}
        p(\bs{V}) = \prod_{k=1}^{K} \prod_{j=1}^{J} \underset{f_{v_{k,j}}}{\underbrace{p(v_{k,j}) }} =\prod_{k=1}^{K} \prod_{j=1}^{J}\mr{Ga} (v_{k,j};\epsilon_{k,j},\eta _{k,j}), 
\end{equation}
where the initial values of $v_{k,j}$ controlled by $\epsilon _{k,j},\eta_{k,j}$. 

\subsection{Factor Graph Representation} 
\begin{figure*}[t]
\centering
    \includegraphics[scale=0.65]{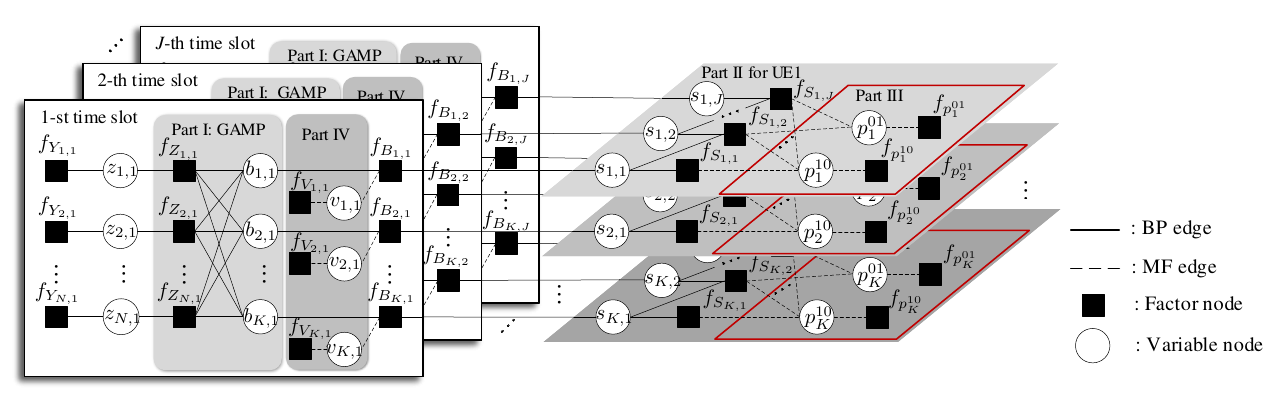}
    \caption{Factor graph representation for the \gls{GF-NOMA} detectors.}
    \label{fig:systemfactor}
\end{figure*}

Based on \eqref{eq:MMV}, \eqref{eq:Likelihood}-\eqref{eq:PrecisionGamma}, the joint \gls{PDF} of all unknown variables given by the observation $\bs{Y}$ can be factorized as 
\begin{align}
    \label{eq:factorize}
    &p(\bs{B},\bs{Z},\bs{S},\bs{V},\bs{p}^{10},\bs{p}^{01} |\bs{Y})\nonumber
        \\
        & = p(\bs{Y}|\bs{Z})p(\bs{Z}|\bs{B})p(\bs{B}|\bs{S},\bs{V},\bs{p}^{10},\bs{p}^{01})\nonumber
        \\
        & \propto \prod_{j=1}^{J} \Big\{ \prod_{n=1}^{N}f_{Y,n,j}(y_{n,j},z_{n,j}) f_{Z,n,j}(z_{n,j},\bs{b}_j) \prod_{k=1}^{K}\nonumber
        \\
        & f_{B_{k,j}}(b_{k,j},s_{k,j},v_{k,j})f_{v_{k,j}}(v_{k,j}) \Big\} \prod_{k=1}^{K} \Big\{ f_{p^{01}_k}(p^{01}_k) f_{p^{10}_k}(p^{10}_k)\nonumber
        \\
        &f_{S_{k,1}}(s_{k,1},p^{10}_{k}) \prod_{j=2}^{J}f_{S_{k,j}}(s_{k,j},s_{k,j-1},p^{10}_{k},p^{01}_{k})\Big\}.
\end{align}
The factor graph of \eqref{eq:factorize} is illustrated in Fig.~\ref{fig:systemfactor}.
\vspace{-0.2cm}

\section{Adaptive detector based on the HMP rule} 
\label{sec:Algorithm}
In this section, we utilize the efficient \gls{GAMP} algorithm to obtain the likelihood function for the signal $b_{k,j},\forall k,j$. We propose the novel \gls{GAMP-BG-MC} algorithm, which can adaptively update the parameters and get the accurate marginal posterior \gls{PDF}. 

\subsection{ Computation of the HMP for GAMP-BG-MC algorithm}
\renewcommand{\thesubsubsectiondis}{Part~\arabic{subsubsection} -}
The probability model in \eqref{eq:ChaPrior} implicit the mix of linear and non-linear functional relationships, i.e., the product and summation operations among the variables as well as the updating of the Gaussian variance. While the single or combined \gls{MP} rules are unavailable to calculate all the messages directly. Fortunately, the \gls{HMP} rule proposed in~\cite{HMP-TSGM}, can tackle the message computation problems with the mix of linear and nonlinear models. The idea of the \gls{HMP} rule~\cite{HMP-TSGM} is that: the \gls{FG} groups all factor nodes into one set $\mathcal{A}_{\mr{Hybrid}}$, while clusters all edges $\mathcal{E}$ into two sets: $\mathcal{E}_{\mr{BP}}$ and $\mathcal{E}_{\mr{MF}}$, depicted as solid and dashed lines in Fig.~\ref{fig:systemfactor}, respectively. For $\mathcal{E}_{\mr{BP}}$ edges, we use the equations \cite[eqs. (5), (6)]{HMP-TSGM} to calculate the messages. Consequently, \cite[eqs. (7), (8) and (9)]{HMP-TSGM} are used to calculate the messages on the $\mathcal{E}_{\mr{MF}}$. For the convenience of \gls{MP} algorithm design, we divide the factor graph into four parts as shown in Fig.~\ref{fig:systemfactor}. 
Then, we will derive the message separately.

\subsubsection{GAMP part messages}
The \gls{GAMP} algorithm is used to obtain the likelihood messages $n^{\mr{BP}}_{b_{k,j}\to f_{B_{k,j}}}(b_{k,j})=\mathcal{CN}(b_{k,j};\hat{q}_{k,j},v_{q_{k,j}})$ of the variable $b_{k,j}$. The detailed scheme of the \gls{GAMP} algorithm can be found in~\cite[Algorithm 1]{MoGAMP}. We first initialize the beliefs $b(v_{k,j})=\mr{Ga}(v_{k,j};\hat{\epsilon}_{k,j},\hat{\eta}_{k,j})$, whose parameters are updated in~\eqref{eq:vparameters}. Then, $m^{\mr{BP}}_{f_{B_{k,j}}\to s_{k,j}}(s_{k,j})$ can be calculated by the \gls{HMP} rule \cite[eq. (5)]{HMP-TSGM}, resulting in
\begin{equation} 
    \label{eq:Sleft}
    m^{\mr{BP}}_{f_{B_{k,j}}\to s_{k,j}}(s_{k,j}) = {\rm{\mathord{\buildrel{\lower3pt\hbox{$\scriptscriptstyle\rightharpoonup$}} 
    \over \pi } }}_{k,j}\delta (1-s_{k,j})+
    (1-{\rm{\mathord{\buildrel{\lower3pt\hbox{$\scriptscriptstyle\rightharpoonup$}} 
    \over \pi } }}_{k,j})\delta (s_{k,j}), 
\end{equation} 
where
\begin{equation}
    \label{eq:phiright}  
    \begin{cases}
        
{\rm{\mathord{\buildrel{\lower3pt\hbox{$\scriptscriptstyle\rightharpoonup$}} 
    \over \pi}}}_{k,j} =\frac{temp_{k,j}}{temp_{k,j}+1}, 
    \\
    temp_{k,j}\triangleq
    \frac{\mr{e}^{\psi(\hat{\epsilon}_{k,j})}\mathcal{CN}(\hat{q}_{k,j};0,v_{q_{k,j}}+\hat{\eta}_{k,j}/\hat{\epsilon}_{k,j})}
    {\hat{\epsilon}_{k,j}\mathcal{CN}(\hat{q}_{k,j};0,v_{q_{k,j}})},
    \\
    \psi(x)\triangleq \mr{ln}x- \frac{1}{2x}
    \end{cases}
\end{equation}

\subsubsection{MC part messages}
Firstly, we initialize the beliefs $b(p^{10}_k)=\mr{Be}(p^{10}_k;\hat{e}_k,\hat{f}_k)$ and $b(p^{01}_k)=\mr{Be}(p^{01}_k;\hat{c}_k,\hat{d}_k)$, the parameters are updated later in \eqref{eq:updateefcd}. The \gls{MC} downward messages are obtained by \gls{HMP} rule \cite[eqs. (5), (6)]{HMP-TSGM} as
\begin{eqnarray} 
    \label{eq:downMSn}
    m^{\mr{BP}}_{f_{S_{k,j}}\to s_{k,j}}(s_{k,j}) =\xi ^\downarrow _{k,j}\delta (1-s_{k,j})+(1-\xi ^\downarrow _{k,j})\delta (s_{k,j}),
     \\
     \label{eq:downNSn}
    n^{\mathrm{BP}}_{s_{k,j}\to f_{S_{k,j+1}}}(s_{k,j})=\xi ^\Downarrow  _{k,j}\delta (1-s_{k,j})+(1-\xi ^\Downarrow  _{k,j})\delta (s_{k,j}),
\end{eqnarray}
where 
\begin{equation}
\label{eq:lambdadown}
    \begin{cases}
    \xi ^\downarrow _{k,1}\triangleq\frac{\mr{exp}\big\{\left \langle \mr{ln}p^{10}_k \right \rangle \big\}}{\mr{exp}\big\{\left \langle  \mr{ln}p^{10}_k\right \rangle \big\}+\mr{exp}\big\{\left \langle \mr{ln}(1-p^{10}_k) \right \rangle \big\}},
    \\
    \xi ^\downarrow _{k,j}\triangleq\frac{\xi ^\Downarrow_{k,j-1} \mathrm{c_1} +
     (1-\xi ^\Downarrow_{k,j-1})
    \mr{c_2}}{\xi ^\Downarrow_{k,j-1} (\mr{c_1}+\mr{c_4})+(1-\xi ^\Downarrow_{k,j-1})(\mr{c_2}+\mr{c_3})},
    \\
    
    \xi^\Downarrow _{k,j}\triangleq\frac{\xi ^\downarrow _{k,j}{\rm{\mathord{\buildrel{\lower3pt\hbox{$\scriptscriptstyle\rightharpoonup$}} \over \pi}}}_{k,j}}
    {\xi ^\downarrow _{k,j}{\rm{\mathord{\buildrel{\lower3pt\hbox{$\scriptscriptstyle\rightharpoonup$}} \over \pi}}}_{k,j}  +
    (1-\xi ^\downarrow _{k,j})(1-{\rm{\mathord{\buildrel{\lower3pt\hbox{$\scriptscriptstyle\rightharpoonup$}} \over \pi}}}_{k,j})},
    \end{cases}
\end{equation}
here, $\mr{c_1},\mr{c_2},\mr{c_3},\mr{c_4}$ denote the constants and defined as
\begin{equation}
    \label{eq:constant}
    \begin{cases}
        \mr{c_1} = \mr{exp}\{\left \langle \mr{ln}(1-p^{01}_k) \right \rangle \} ;~\mr{c_2} = \mr{exp}\{\left \langle \mr{ln}p^{10}_k \right \rangle \};  
        \\
        \mr{c_3} = \mr{exp}\{\left \langle \mr{ln}(1-p^{10}_k) \right \rangle \};~\mr{c_4} = \mr{exp}\{\left \langle \mr{ln}p^{01}_k \right \rangle \}.
    \end{cases}
\end{equation}
where $\left \langle \mr{ln}x \right \rangle _{\mr{Be}(x;a,b)}=\psi (a)-\psi (a+b), \left \langle \mr{ln}(1-x) \right \rangle _{\mr{Be}(x;a,b)}=\psi (b)-\psi (a+b)$.

Similarly, the \gls{MC} upward messages passed between ${s_{k,j}}$ and ${f_{S_{k,j}}}$ can be computed by~\cite[eqs. (5), (6)]{HMP-TSGM}, obtaining
\begin{eqnarray}
    \label{eq:upNSn}
    n^{\mr{BP}}_{s_{k,j}\to f_{S_{k,j}}}(s_{k,j})\!=\!\xi^\Uparrow_{k,j}\delta(1-s_{k,j})+(1-\xi^\Uparrow_{k,j})\delta(s_{k,j}),
    \\
    \label{eq:upMSn}
    m^{\mr{BP}}_{f_{S_{k,j+1}}\to s_{k,j}}(s_{k,j}) \!=\!
    \xi ^\uparrow _{k,j}\delta (1-s_{k,j})+(1-\xi ^\uparrow _{k,j})\delta (s_{k,j}),
\end{eqnarray} 
where
\begin{equation}
\label{eq:lambdaup}
    \begin{cases}
    \xi^\Uparrow_{k,J}\triangleq{\rm{\mathord{\buildrel{\lower3pt\hbox{$\scriptscriptstyle\rightharpoonup$}} \over \pi}}}_{k,j},~~
    \xi^\Uparrow_{k,j} \triangleq \frac{\xi ^\uparrow _{k,j}{\rm{\mathord{\buildrel{\lower3pt\hbox{$\scriptscriptstyle\rightharpoonup$}} \over \pi}}}_{k,j}}
    {\xi ^\uparrow _{k,j}{\rm{\mathord{\buildrel{\lower3pt\hbox{$\scriptscriptstyle\rightharpoonup$}} \over \pi}}}_{k,j}+
    (1-\xi ^\uparrow _{k,j})(1-{\rm{\mathord{\buildrel{\lower3pt\hbox{$\scriptscriptstyle\rightharpoonup$}} \over \pi}}}_{k,j})},
    \\
    \xi ^\uparrow _{k,j} \triangleq \frac{\xi^\Uparrow_{k,j+1}\mr{c_1}+(1-\xi^\Uparrow_{k,j+1}) \mr{c_4}}
    {\xi ^\Uparrow_{k,j+1} (\mr{c_1}+\mr{c_2})+(1-\xi^\Uparrow_{k,j+1})(\mr{c_3}+\mr{c_4})},    
    \end{cases}
\end{equation}
here $\mr{c_1},\mr{c_2},\mr{c_3},\mr{c_4}$ are the same as \eqref{eq:constant}.

\subsubsection{Transition probability update part}
Given the messages $p(p^{10}_k)$ and $p(p^{01}_k)$, we can update the MC transition probability parameters using the method in \cite[Part 4 within Section 4]{HMP-TSGM}.the beliefs update by  $b(p^{10}_k)\propto \mr{Be}(p^{10}_k;\hat{e}_k,\hat{f}_k)$ and $b(p^{01}_k)\propto \mr{Be}(p^{01}_k;\hat{c}_k,\hat{d}_k)$, where the parameters are updated through
\begin{equation}
    \label{eq:updateefcd}
    \begin{cases}
    \hat{e}_k=B_{s_{k,1}}\!+e_k+\!\sum\limits_{j=2}^{J} \frac{B_{k,j}^{10}}{\rho_{k,j}},\quad \hat{c}_k=c_k\!+\!\sum\limits_{j=2}^{J} \frac{B_{k,j}^{01}}{\rho_{k,j}},
    \\
    \hat{f}_k=1\!-\!B_{s_{k,1}}\!+\!f_k\!+\!\sum\limits_{j=2}^{J}\frac{B_{k,j}^{00}}{\rho_{k,j}},\hat{d}_k=d_k\!+\!\sum\limits_{j=2}^{J}\frac{B_{k,j}^{11}}{\rho_{k,j}}.
    \end{cases}
\end{equation}
and the parameters $B_{s_{k,1}},B_{k,j}^{00},B_{k,j}^{01},B_{k,j}^{10},B_{k,j}^{11},\rho_{k,j}$ can be calculated by \cite[eq. (40)]{HMP-TSGM}.

\subsubsection{Variance update part}
The message going out of the \gls{MC} from $s_{k,j}$ to $f_{B_{k,j}}$ can be computed by \cite[eq. (6)]{HMP-TSGM}
\begin{equation}
    \label{eq:Sright}
    n^{\mr{BP}}_{s_{k,j}\to f_{B_{k,j}}}(s_{k,j})={\rm{\mathord{\buildrel{\lower3pt\hbox{$\scriptscriptstyle\leftharpoonup$}}\over\pi}}}_{k,j}\delta(1-s_{k,j})+(1-{\rm{\mathord{\buildrel{\lower3pt\hbox{$\scriptscriptstyle\leftharpoonup$}}\over\pi}}}_{k,j})\delta(s_{k,j}), 
\end{equation}
where
\begin{equation}
    \label{eq:phileft}
    {\rm{\mathord{\buildrel{\lower3pt\hbox{$\scriptscriptstyle\leftharpoonup$}} 
    \over \pi}}}_{k,j} \triangleq\frac{\xi^\uparrow_{k,j}\xi^\downarrow_{k,j}}
    {\xi^\uparrow_{k,j}\xi^\downarrow_{k,j}+
    (1-\xi^\uparrow_{k,j})(1-\xi^\downarrow_{k,j})}.
\end{equation} 

To update the variance of the BG-MC model, we have to calculate the combined belief $b(b_{k,j},s_{k,j})$ using \cite[eq. (8)]{HMP-TSGM} as
\begin{equation}
   \begin{split}
    \label{eq:combands}
    b(b_{k,j},s_{k,j})=(1-B_{b_{k,j},s_{k,j}})\delta(s_{k,j})\delta(b_{k,j})+
    \\
    B_{b_{k,j},s_{k,j}}\delta(1-s_{k,j})\mathcal{CN}(b_{k,j};\hat{\mu}_{k,j},\vartheta_{k,j}),
     \end{split}
\end{equation}
where
\begin{equation}
\label{eq:muandv}
    \begin{cases}
        B_{b_{k,j},s_{k,j}}\triangleq\frac{{\rm{\mathord{\buildrel{\lower3pt\hbox{$\scriptscriptstyle\leftharpoonup$}} \over \pi}}}_{k,j}temp_{k,j}}{{\rm{\mathord{\buildrel{\lower3pt\hbox{$\scriptscriptstyle\leftharpoonup$}} \over \pi}}}_{k,j}temp_{k,j}+(1-{\rm{\mathord{\buildrel{\lower3pt\hbox{$\scriptscriptstyle\leftharpoonup$}} \over \pi}}})},
        \\
        \vartheta_{k,j}=(v^{-1}_{q_{k,j}}+\frac{\hat{\epsilon}_{k,j}}{\hat{\eta}_{k,j}})^{-1},~~\hat{\mu}_{k,j}=\frac{\hat{q}_{k,j}}{v_{q_{k,j}}}\vartheta_{k,j}.
    \end{cases}
\end{equation}
and the temporary variable $temp_{k,j}$ is defined in \eqref{eq:phiright}.

Then, the message from $f_{B_{k,j}}$ to variable node $v_{k,j}$ can be computed with the \gls{HMP} rule~\cite[eq. (7)]{HMP-TSGM} as
\begin{equation}
  \begin{split}
     \label{eq:Btov}
    &m^{\mr{MF}}_{f_{B_{k,j}}\to v_{k,j}}(v_{k,j})
    \\
    &\propto\mr{Ga}(v_{k,j};B_{b_{k,j},s_{k,j}}+1,B_{b_{k,j},s_{k,j}}(\lvert\hat{\mu}_{k,j}\rvert^2+\vartheta_{k,j})).
  \end{split}
\end{equation}
Given $m^{\mr{MF}}_{f_{v_{k,j}}\to v_{k,j}}(v_{k,j})=\mr{Ga}(v_{k,j};\epsilon_{k,j},\eta_{k,j})$, we can update the belief as $b(v_{k,j})=\mr{Ga}(v_{k,j};\hat{\epsilon}_{k,j},\hat{\eta}_{k,j})$, where parameters are updated through
\begin{equation}
    \begin{cases}
        \label{eq:vparameters}
        \hat{\epsilon}_{k,j}=\epsilon_{k,j}+B_{b_{k,j},s_{k,j}},
        \\
        \hat{\eta}_{k,j}=\eta_{k,j}+B_{b_{k,j},s_{k,j}}(\lvert\hat{\mu}_{k,j}\rvert^2+\vartheta_{k,j}).
    \end{cases}
\end{equation}
With the parameters updated in \eqref{eq:vparameters}, we can compute the message from $f_{B_{k,j}}$ to $b_{k,j}$ by HMP rule~\cite[eq. (5)]{HMP-TSGM}.
Finally, we can get the belief $b(b_{k,j})=(1-B_{b_{k,j}})\delta(b_{k,j})+B_{b_{k,j}}\mathcal{CN}(b_{k,j};\hat{\mu}_{k,j},\vartheta_{k,j})$. 
with the parameter $B_{b_{k,j}}=B_{b_{k,j},s_{k,j}}$. Then, the algorithm outputs the mean and variance of the belief $b(b_{k,j})$ as
\begin{equation}
   \begin{cases}
        \label{eq:meanandvariance}
        \hat{b}_{k,j}=\left\langle b_{k,j}\right\rangle_{b(b_{k,j})}=B_{b_{k,j}}\hat{\mu}_{k,j},
        \\
        v_{b_{k,j}}=B_{b_{k,j}}(\lvert\hat{\mu}_{k,j}\rvert^2+\vartheta_{k,j})-\lvert \hat{b}_{k,j}\rvert^2.
   \end{cases}
\end{equation}
\begin{algorithm}[ht]
    \DontPrintSemicolon
    \SetNoFillComment
    
    \caption{\small{GAMP-BG-MC algorithm}}
    \label{alg:Hybrid}
    \footnotesize
    \KwInput{Received signal $\bs{Y}$, channel matrix $\bs{H}$, iteration number $T$.}
    \KwOutput{$\mr{Decoding~results~} \hat{b}_{k,j},\forall k,j.$}
    \SetKwInput{KwInitialize}{Initialize}
    \KwInitialize{Prior parameters: $\epsilon_{k,j},\eta_{k,j},\forall k,j; e_k,f_k,c_k,d_k,\forall k.$
    \\
    Belief parameters: $\hat{\epsilon}_{k,j},\hat{\eta}_{k,j},\forall k,j;\hat{e}_k,\hat{f}_k,\hat{c}_k,\hat{d}_k,\xi^{\Uparrow }_{k,J},\forall k$.}
      
    \While{ convergence == FALSE OR iteration number $<$ T}{
    
    \tcp*[l]{Part 1 - GAMP part messages}
    Run \cite[Algorithm 1]{MoGAMP} and output $v_{q_{k,j}},\hat{q}_{k,j}, \forall k,j$\;
    
    $\forall k,j$: update ${\rm{\mathord{\buildrel{\lower3pt\hbox{$\scriptscriptstyle\rightharpoonup$}} \over \pi}}}_{k,j}$ by \eqref{eq:phiright}\;
    
    \tcp*[l]{Part 2 - MC part messages}
    $\forall k$: update $\xi^{\downarrow }_{k,1}$ by \eqref{eq:lambdadown}\; 
   
    $\forall k,j\in[2:J]$: update $\xi^{\downarrow }_{k,j}$ by \eqref{eq:lambdadown}\;

    $\forall k,j$: update $\xi^{\Downarrow }_{k,j}$ by \eqref{eq:lambdadown}\;
    
    $\forall k,j\in[J-1:1]$: update $\xi^{\Uparrow }_{k,j}$ and $\xi^{\uparrow }_{k,j}$ by \eqref{eq:lambdaup}\; 

    \tcp*[l]{Part 3 - Transition probability update part}
    
    $\forall k,j$: update $B_{s_{k,1}},B_{k,j}^{00},B_{k,j}^{01},B_{k,j}^{10},B_{k,j}^{11},\rho_{k,j}$ by (40) in \cite{HMP-TSGM}\;
    
    $\forall k$: update $\hat{e}_k,\hat{f}_k~\mr{and}~\hat{c}_k,\hat{d}_k$ by \eqref{eq:updateefcd}\;
    
    perform lines 4-7 again\;
    
    \tcp*[l]{Part 4 - Variance update part}
    $\forall k,j$: update ${\rm{\mathord{\buildrel{\lower3pt\hbox{$\scriptscriptstyle\leftharpoonup$}} \over \pi}}}_{k,j}$ by \eqref{eq:phileft}\;

    $\forall k,j$: update $B_{b_{k,j},s_{k,j}}$ and $\hat{\mu}_{k,j},\vartheta_{k,j}$ by \eqref{eq:muandv}\;

    $\forall k,j$: update $\hat{\epsilon}_{k,j},\hat{\eta}_{k,j}$ by \eqref{eq:vparameters}\;

    $\forall k,j$: update $B_{b_{k,j}}$ and $\hat{\mu}_{k,j},\vartheta_{k,j}$ by \eqref{eq:muandv} again based on new updated line 13\;

    $\forall k,j$: update $\hat{b}_{k,j},v_{b_{k,j}}$ by \eqref{eq:meanandvariance}\;
    }  
\end{algorithm}
\vspace{-0.5cm}

\subsection{Message Passing Scheduling}
We summarize the proposed \gls{GAMP-BG-MC} algorithm as shown in \textbf{Algorithm~\ref{alg:Hybrid}}. Specifically, the \emph{Part~1} messages are first calculated in parallel at each time slot, which contains the~\cite[Algorithm 1]{MoGAMP} and the message going into the \emph{Part~2}. Then we calculate the \gls{MC} messages in sequence in lines 4-7. In \emph{Part~3}, the parameters of $p^{10}_{k},p^{01}_{k}$ can be updated in lines 8-10. After that, we can improve the accuracy of the outgoing messages in \emph{Part~4} by updating the \gls{MC} part messages again based on the updated hyperparameters $p^{10}_k,p^{01}_k$. In the final step, we update the parameters of variances and calculate the output estimated data $\hat{b}_{k,j},\forall k,j$ with lines 12-15 in \emph{Part~4}. 
\renewcommand{\thesubsubsectiondis}{\arabic{subsubsection})}
\begin{figure*}[th]
\centering
    \begin{subfigure}[t]{.3\textwidth}
        \centering
        \includegraphics[scale=0.31]{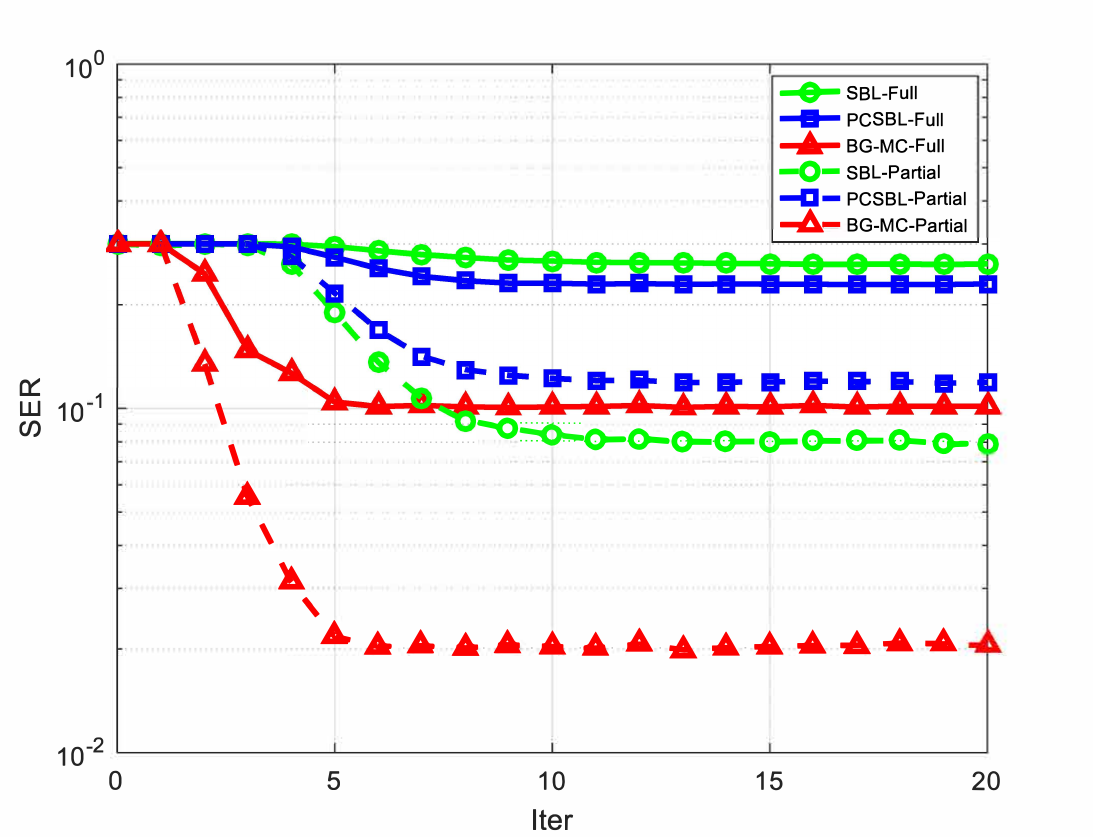}
        \caption{SER vs Iteration with $p_a = 0.3$.}
        \label{fig:SERvsIter}
    \end{subfigure}
    \hfill
    \begin{subfigure}[t]{.3\textwidth}
        \centering
        \includegraphics[scale=0.31]{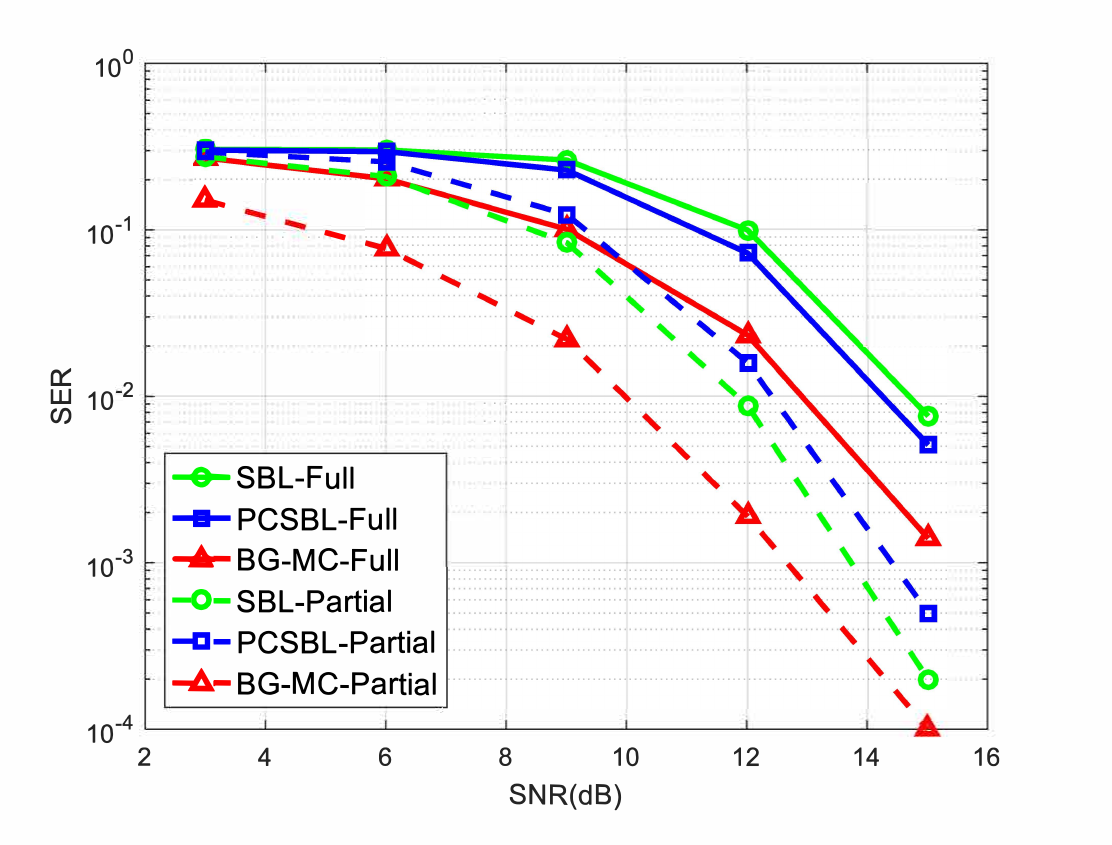}
        \caption{SER vs SNR with $p_a = 0.3.$}
        \label{fig:SERvsSNR}
    \end{subfigure}
    \hfill
    \begin{subfigure}[t]{.3\textwidth}
        \centering
        \includegraphics[scale=0.31]{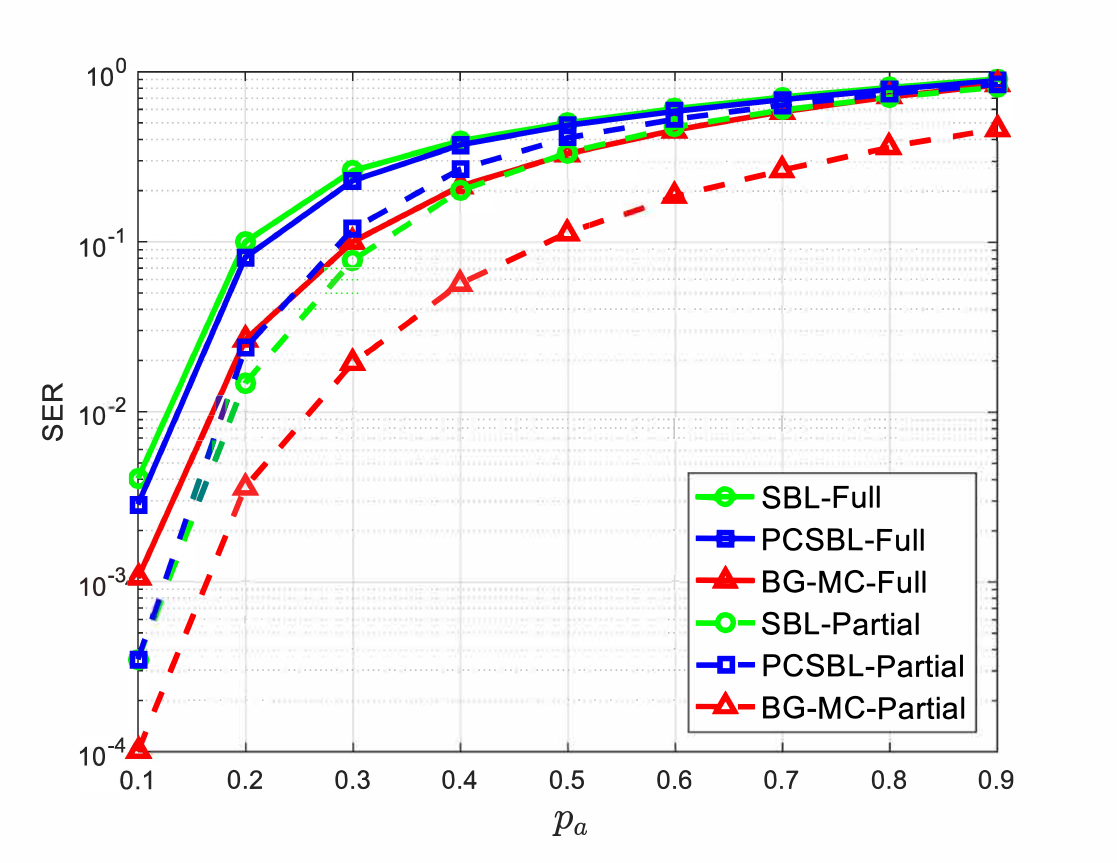}
        \caption{SER vs $p_a$ with $\mr{SNR} = 9$dB.}
        \label{fig:SERvspa}
    \end{subfigure}
\caption{Performance of proposed GAMP-BG-MC algorithm.}
\label{fig:simulation}
\end{figure*}
\section{Simulation Results}
\label{Sec:simulation_results}
In this section, we compare the performance of the proposed GAMP-BG-MC algorithm. As a benchmark, we consider the state-of-the-art algorithms \gls{GAMP-SBL}~\cite{ZhangPCSBLTVT,ZhangPCSBLTWC}, \gls{GAMP-PCSBL}~\cite{ZhangPCSBLTVT} referred as \gls{SBL} and \gls{PCSBL} in the following. Referring to \cite{ZhangPCSBLTVT}, we consider the BPSK modulation scheme in all simulations, the \gls{UE} number $K=20$, the spreading factor $N = 30$, and the number of the continuous time slots $J = 6$. In experiments, we design two scenarios compared with the benchmark algorithms, one is the \gls{UE} activity with temporal correlation, i.e., the \gls{UE} will transmit the data partially during the whole frame, here denoted as "Partial". The other is an extreme phenomenon, the \gls{UE} keeps transmitting data fully to the frame, referred to as "Full". 

For the initialization, according to \cite{HMP-TSGM}, the prior parameters are $\{\epsilon_{k,j}=\eta_{k,j}=1,\forall k,j\}$, $\{e_k=f_k=c_k=d_k=1,\forall k\}$, $\xi^{\Uparrow }_{k,j}\!=\!1/2,\forall k,j$, and the belief parameters are initialized with the same values of the priors. All the experiments are obtained by averaging $\mr{sim} = 1000$ Monte Carlo realizations. We use the \gls{SER} defined as $ \mr{SER} =\mr{log}_{10}(\frac{K_{err}}{\mr{sim}\times K})$ as a performance metric, $K_{err}$ is the number of \gls{UE} whose symbol is erroneously decoded. The per-iteration complexity of the \gls{BG-MC} is $\mathcal{O}((N+1)KJ)$, and it is in the same order as the SBL and PCSBL. But the iteration number of the \gls{BG-MC} will always be small, as shown in Fig.~\ref{fig:simulation} (a).

Fig.~\ref{fig:simulation} (a) gives the convergence speed of the three algorithms under the \gls{SNR} = 9dB. The proposed BG-MC algorithm has the fastest convergence speed compared with the SBL and PCSBL algorithms in both two scenarios. In Fig.~\ref{fig:simulation} (b), we compare the average \gls{SER} performance as a function of the \glspl{SNR}. The simulation results show that the BG-MC algorithm performs better in the two scenarios. Further, we can find that the PCSBL is better than the SBL in the second scheme and the PCSBL is efficient in characterizing the block sparse structure, while the \gls{BG-MC} is more accurate not only in the block sparse structure but also in the cluster sparse structure. The reason is that the \gls{BG-MC} can update the statistic parameters automatically, which is suitable for the practice system. Fig.~\ref{fig:simulation} (c) shows the \gls{SER} performances as a function of the \gls{UE} active rate $p_a$ under \gls{SNR} = 9dB. The BG-MC algorithm outperforms the benchmark algorithms in almost all ranges of $p_a$, especially in the first scenario.

\section{Conclusion}
\label{sec:conclusion}
In this paper, we developed the BG-MC model to accurately characterize the slow change feature of the \gls{UE} \gls{AS} under the \gls{GF-NOMA} system. Based on the \gls{HMP} rule, we designed the GAMP-BG-MC algorithm and tested it with the two scenarios. Experiments show that the proposed algorithm achieves better \gls{SER} performance. Hence, the proposed technique can be deployed for the URLLC in the 6G and beyond.

\bibliographystyle{IEEEtran}
\bibliography{IEEEabrv,reference}
\end{document}